\documentclass[a4paper,11pt]{article}
\usepackage{pos}

\newcommand{\bfp}{{\bf p}_{\perp}}

\newcommand{\Dp}{{\bf \Delta}_{\perp}}

\newcommand{\be}{\begin{eqnarray}}
\newcommand{\ee}{\end{eqnarray}}

\title{Study of the subleading twist GTMD $E_{21}$ for proton in light-front quark-diquark model}

\author[a]{Sameer Jain}
\author[a]{Shubham Sharma}
\author*[a]{Harleen Dahiya}

\affiliation[a]{Dr. B. R. Ambedkar National Institute of Technology, Jalandhar 144008, India
}

\emailAdd{sameerjainofficial@gmail.com}
\emailAdd{s.sharma.hep@gmail.com}
\emailAdd{dahiyah@nitj.ac.in}

\abstract{In this work, we study the generalized transverse momentum dependent distribution (GTMD) $E_{21}$ for proton using light-front quark-diquark model. We construct the expression of $E_{21}$ GTMD using the overlap equation in light-front wave functions obtained from the GTMD correlator with Dirac matrix structure $\Gamma=1$, in both situations of scalar and vector diquark. The $3$-dimensional plots of GTMD $E_{21}$ have been analyzed with respect to its variables by taking two variables at a time while holding others constant.}

\FullConference{The XVIth Quark Confinement and the Hadron Spectrum Conference (QCHSC24)\\
 19-24 August, 2024\\
 Cairns Convention Centre, Cairns, Queensland, Australia\\}

\begin{document}
\maketitle

\section{Introduction \label{secintro}}
\noindent
Generalized transverse momentum dependent distributions (GTMDs), often referred to as the ``mother distributions,'' are indispensable tools for unraveling the multidimensional structure of hadrons. Their intricate mathematical framework encapsulates the correlations between parton transverse momentum distributions and spatial distributions, offering a comprehensive picture of hadronic structure \cite{Meissner:2009ww, Lorce:2013pza, Meissner:2008ay, Echevarria:2016yza}. These distributions provide a theoretical basis for investigating phenomena such as orbital angular momentum and spin-orbit coupling, which are key to understanding the spin dynamics of hadrons \cite{Lorce:2011ni, Hatta:2011ku}. Furthermore, GTMDs serve as an important bridge between theoretical predictions and experimental observables, by connecting to both transverse momentum-dependent distributions (TMDs) and generalized parton distributions (GPDs).
\par
While the leading-twist GTMDs have been extensively studied due to their direct relevance to experimental measurements, the subleading twist counterparts remain relatively unexplored \cite{Maji:2022tog, Bhattacharya:2017bvs}. These higher-twist contributions, however, play a vital role in revealing the full intricacies of hadronic dynamics, including quark-gluon correlations, intrinsic transverse motion, and higher-order interactions \cite{Sharma:2024arf}. Addressing these subleading twist effects is essential for a comprehensive understanding of the non-perturbative aspects of Quantum Chromodynamics (QCD).
\par
To explore these dynamics, we employ the light-front quark-diquark model (LFQDM), a framework that conceptualizes the proton as a bound state of an active quark and a diquark spectator \cite{Maji:2016yqo}. This model, known for its simplicity and effectiveness, has been widely utilized to capture non-perturbative features of hadronic structure \cite{Maji:2016yqo, Maji:2017ill, Maji:2017bcz}. 
In this work, we focus on computing the subleading twist GTMD $E_{2,1}$, analyzing its physical significance, and exploring its connection to experimental observables.
\par
The present study offers fresh perspectives on the role of subleading twist GTMDs in the broader context of hadron structure. Our findings contribute to a deeper understanding of the interplay between parton-level dynamics and spin-dependent effects, providing valuable insights for theoretical developments and experimental efforts at facilities such as the upcoming Electron-Ion Colliders (EICs).
\section{Light-Front Quark-Diquark Model  \label{secmodel}}
\noindent 
The LFQDM \cite{Maji:2016yqo} conceptualizes the proton as a composite system, where an active quark interacts dynamically with the external probe, while the diquark remains a spectator \cite{Chakrabarti:2019wjx}. The proton’s spin-flavor structure follows $SU(4)$ symmetry and consists of an isoscalar-scalar diquark singlet $|u~ S^0\rangle$, an isoscalar-vector diquark $|u~ A^0\rangle$, and an isovector-vector diquark $|d~ A^1\rangle$. The proton state is then represented as 
\be
|P; \pm\rangle = C_S |u~ S^0\rangle^\pm + C_V |u~ A^0\rangle^\pm + C_{VV} |d~ A^1\rangle^\pm \,,
\ee
as discussed in Ref. \cite{Jakob:1997wg, Bacchetta:2008af}. In the above equation, $S$ and $A = V, VV$ denote the scalar and vector diquarks, with the superscripts indicating their isospin characteristics.
Both scalar and vector diquark coefficients $C_{i}$ are provided in Ref. \cite{Maji:2016yqo} and are given as
		$C_{S}^{2} =1.3872$, 
		$C_{V}^{2} =0.6128$, 
		 and $C_{V V}^{2} =1$.

%
 The Fock-state expansion for the case of a scalar diquark with the angular momentum component $J^z = \pm 1/2$ can be expressed as \cite{majiref25}
%
\begin{eqnarray}
		|u~ S\rangle^\pm &=&\sum_{\lambda^q}  \int \frac{dx~ d^2\bfp}{2(2\pi)^3\sqrt{x(1-x)}}  \psi^{\pm(\nu)}_{\lambda^q}\left(x,\bfp\right)\bigg|\lambda^{q},\lambda^{S}; xP^+,\bfp\bigg\rangle .\label{fockSD}
\end{eqnarray}
Accordingly, for the case of a vector diquark, the state is
	\begin{eqnarray}
		|\nu~ A \rangle^\pm &=&\sum_{\lambda^q} \sum_{\lambda^D} \int \frac{dx~ d^2\bfp}{2(2\pi)^3\sqrt{x(1-x)}}  \psi^{\pm(\nu)}_{\lambda^q \lambda^D }\left(x,\bfp\right)\bigg|\lambda^{q},\lambda^{D}; xP^+,\bfp\bigg\rangle ,\label{fockVD}
	\end{eqnarray}

Here, $|\lambda^q~\lambda^S; xP^+,\bfp\rangle$ ($|\lambda^q~\lambda^D; xP^+,\bfp\rangle$) represents the state of two particles, where the helicity of a active quark is $\lambda^q$ and the helicity of a scalar diquark is $\lambda^S=s$ (singlet) and for vector diquark  it is $\lambda^D=\pm 1,0$ (triplet). The flavor index is represented by $\nu~(=u,d)$.
The LFWFs for the scalar diquark and vector diquark are provided in Table \ref{tab_LFWF_s} and \ref{tab_LFWF_v} accordingly. as per Ref. \cite{Maji:2017bcz}.
\begin{table}[h]
\centering 
\begin{tabular}{|p{0.7cm}|p{0.67cm}|p{1.6cm}p{0.05cm} p{2.8cm}|p{1.6cm}p{0.05cm}p{2.8cm}|}
\hline
~~$\lambda^q$~~&~~$\lambda^S$~~&\multicolumn{3}{c|}{LFWFs for $J^z=+1/2$} & \multicolumn{3}{c|}{LFWFs for $J^z=-1/2$}\\
\hline
$+1/2$~~&~~$~0$~~&$\psi^{+(\nu)}_{+}(x,\bfp)$&~~=&$~N_S~ \varphi^{(\nu)}_{1}$~~&$\psi^{-(\nu)}_{+}(x,\bfp)$~&~~=&$~N_S \bigg(\frac{p^1-ip^2}{xM}\bigg)~\varphi^{(\nu)}_{2}$~~\\
$-1/2$~~&~~$~0$~~&$\psi^{+(\nu)}_{-}(x,\bfp)$&~~=&$~-N_S\bigg(\frac{p^1+ip^2}{xM} \bigg)~ \varphi^{(\nu)}_{2}$&$\psi^{-(\nu)}_{-}(x,\bfp)$&~~=&$~N_S~ \varphi^{(\nu)}_{1}$~~\\
\hline
\end{tabular}
\caption{The scalar diquark LFWFs for the scenario when $J^z=\pm1/2$, for various values of the active quark helicities $\lambda^q$ and $\lambda^S=0$, where $N_S$ is the normalization constant.}
\label{tab_LFWF_s} 
\end{table}
\begin{table}[h]
\centering 
\begin{tabular}{ |p{0.78cm}|p{0.68cm}|p{1.5cm}p{0.05cm}p{3.4cm}|p{1.8cm}p{0.05cm}p{3.5cm}|}
\hline
~~~$\lambda^q$~~&~~$\lambda^D$~~&\multicolumn{3}{c|}{LFWFs for $J^z=+1/2$} & \multicolumn{3}{c|}{LFWFs for $J^z=-1/2$}\\
\hline
~$+1/2$~~&~~$+1$~~&$\psi^{+(\nu)}_{+~+}(x,\bfp)$&~~=&$~~N^{(\nu)}_1 \sqrt{\frac{2}{3}} \bigg(\frac{p^1-ip^2}{xM}\bigg)~  \varphi^{(\nu)}_{2}$~~&~~$\psi^{-(\nu)}_{+~+}(x,\bfp)$&~~=&$~~0$~~  \\
~$-1/2$~~&~~$+1$~~&$\psi^{+(\nu)}_{-~+}(x,\bfp)$~&~~=&$~~N^{(\nu)}_1 \sqrt{\frac{2}{3}}~ \varphi^{(\nu)}_{1}$~~&~~$\psi^{-(\nu)}_{-~+}(x,\bfp)$~&~~=&$~~0$~~   \\
~$+1/2$~~&~~$~~0$~~&$\psi^{+(\nu)}_{+~0}(x,\bfp)$~&~~=&$~~-N^{(\nu)}_0 \sqrt{\frac{1}{3}}~  \varphi^{(\nu)}_{1}$~~&~~$\psi^{-(\nu)}_{+~0}(x,\bfp)$~&~~=&$~~N^{(\nu)}_0 \sqrt{\frac{1}{3}} \bigg( \frac{p^1-ip^2}{xM} \bigg)~  \varphi^{(\nu)}_{2}$~~   \\
~$-1/2$~~&~~$~~0$~~&$\psi^{+(\nu)}_{-~0}(x,\bfp)$~&~~=&$~~N^{(\nu)}_0 \sqrt{\frac{1}{3}} \bigg(\frac{p^1+ip^2}{xM} \bigg)~ \varphi^{(\nu)}_{2}$~~&~~$\psi^{-(\nu)}_{-~0}(x,\bfp)$~&~~=&$~~N^{(\nu)}_0\sqrt{\frac{1}{3}}~  \varphi^{(\nu)}_{1}$~~   \\
~$+1/2$~~&~~$-1$~~&$\psi^{+(\nu)}_{+~-}(x,\bfp)$~&~~=&$~~0$~~&~~$\psi^{-(\nu)}_{+~-}(x,\bfp)$~&~~=&$~~- N^{(\nu)}_1 \sqrt{\frac{2}{3}}~  \varphi^{(\nu)}_{1}$~~   \\
~$-1/2$~~&~~$-1$~~&$\psi^{+(\nu)}_{-~-}(x,\bfp)$~&~~=&$~~0$~~&~~$\psi^{-(\nu)}_{-~-}(x,\bfp)$~&~~=&$~~N^{(\nu)}_1 \sqrt{\frac{2}{3}} \bigg(\frac{p^1+ip^2}{xM}\bigg)~  \varphi^{(\nu)}_{2}$~~   \\
\hline
\end{tabular}
\caption{The LFWFs for the vector diquark for the case when $J^z=\pm1/2$, for different values of helicities of active quark $\lambda^q$ and vector diquark $\lambda^D$. $N^{(\nu)}_0$, $N^{(\nu)}_1$ are the normalization constants.}
\label{tab_LFWF_v} 
\end{table}
In Tables \ref{tab_LFWF_s} and \ref{tab_LFWF_v}, the generic ansatz of LFWFs $\varphi^{(\nu)}_{i}=\varphi^{(\nu)}_{i}(x,\bfp)$ is based on the soft-wall AdS/QCD prediction \cite{BT,majiref27}, and the parameters $a^\nu_i,~b^\nu_i$, and $\delta^\nu$ have been determined as \cite{Maji:2017bcz}
\begin{equation}
\varphi_i^{(\nu)}(x,\bfp)=\frac{4\pi}{\kappa}\sqrt{\frac{\log(1/x)}{1-x}}x^{a_i^\nu}(1-x)^{b_i^\nu}\exp\Bigg[-\delta^\nu\frac{\bfp^2}{2\kappa^2}\frac{\log(1/x)}{(1-x)^2}\bigg].
\label{LFWF_phi}
\end{equation}
Using the Dirac and Pauli form factor data, the parameters $a_i^{\nu}$ and $b_i^{\nu}$ that appear in Eq. \eqref{LFWF_phi} have been fitted to the model scale $\mu_0=0.313{\ \rm GeV}$ \cite{Maji:2016yqo,majiref16,majiref17}. Table \ref{tab_par} has a list of all of the relevant parameters.
\begin{table}[h]
\centering 
\begin{tabular}{|p{0.5cm}|c|c|c|c|p{0.5cm}|}
\hline
~~$\nu$~~&~~$a_1^{\nu}$~~&~~$b_1^{\nu}$~~&~~$a_2^{\nu}$~~&~~$b_2^{\nu}$~~&~~$\delta^{\nu}$\\
\hline
~~$u$~~&~~$0.280\pm 0.001$~~&~~$0.1716 \pm 0.0051$~~&~~$0.84 \pm 0.02$~~&~~$0.2284 \pm 0.0035$~~&~$1.0$\\
~~$d$~~&~~$0.5850 \pm 0.0003$~~&~~$0.7000 \pm 0.0002$~~&~~$0.9434^{+0.0017}_{-0.0013}$~~&~~$0.64^{+0.0082}_{-0.0022}$~~&~$1.0$\\
\hline
\end{tabular}
\caption{
Model parameter values for the $u$ and $d$ quarks for the LFWF expression given in Eq.~\eqref{LFWF_phi}.}
\label{tab_par} 
\end{table}
In addition, the normalization constants $N_{i}^{2}$ in Table {\ref{tab_LFWF_s}} and {\ref{tab_LFWF_v}} are listed in Table \ref{tab_NC} and are calculated in line with Ref. \cite{Maji:2016yqo}.
\begin{table}[h]
\centering 
\begin{tabular}{ |p{0.68cm}|p{1.2cm}|p{1.2cm}|p{1.2cm}|  }
\hline
~$~~\nu$&~~~~$N_{S}$~~&~~$~~N_0^{\nu}$~~&~~$~~N_1^{\nu}$~~  \\
\hline
~$~~u$&~~$2.0191$~~&~~$3.2050$~~&~~$0.9895$~~  \\
~$~~d$&~~$2.0191$~~&~~$5.9423$~~&~~$1.1616$~~    \\
\hline
\end{tabular}
\caption{Values of normalization constants $N_{i}^{2}$ which appears in Table {\ref{tab_LFWF_s}} and {\ref{tab_LFWF_v}}, corresponding to both $u$ and $d$ quark.}
\label{tab_NC} 
\end{table}
The light-cone convention $z^\pm=z^0 \pm z^3$ is used in the calculations, and the frame is chosen so that the average transverse momentum of the proton is zero, i.e., $P \equiv \big(P^+,\frac{M^2}{P^+},\textbf{0}_\perp\big)$. 
The value of $0.4~\mathrm{GeV}$ has been allocated to the parameter $\kappa$ for the AdS/QCD scale, which appears in Eq. (\ref{LFWF_phi}) \cite{majiref28}. The constituent quark mass ($m$) and proton mass ($M$) have been taken to be $0.055~\mathrm{GeV}$ and $0.938~\mathrm{GeV}$, respectively, in accordance with Ref. \cite{Chakrabarti:2019wjx}.

\section{GTMD Correlator}\label{secgtmdcorr}
\noindent 
As stated in Ref.~\cite{Meissner:2009ww}, the fully unintegrated GTMD correlator for a  spin-$\frac{1}{2}$ hadron at fixed light-front time \(z^+ = 0\) is given by
\begin{eqnarray} 
	W^{\nu [\Gamma]}_{[\Lambda^{N_i}\Lambda^{N_f}]}=\frac{1}{2}\int \frac{dz^-}{(2\pi)} \frac{d^2z_T}{(2\pi)^2} e^{ip.z} 
	\langle P^{f}; \Lambda^{N_f} |\bar{\psi} (-z/2)\Gamma \mathcal{W}_{[-z/2,z/2]} \psi (z/2) |P^{i};\Lambda^{N_i}\rangle \bigg|_{z^+=0}\,.
 \label{corr}
\end{eqnarray}
Here, the initial and final states of the proton, with helicities $\Lambda^{N_i}$ and $\Lambda^{N_f}$, are denoted by $|P^{i};\Lambda^{N_i}\rangle$ and $|P^{f}; \Lambda^{N_f}\rangle$, respectively. The operator $\psi$ represents the quark field. The Wilson line $\mathcal{W}_{[-z/2,z/2]}$ ensures $SU(3)$ color gauge invariance of the bilocal quark operator, and for simplicity, we set it to 1. As discussed in Ref.~\cite{Sharma:2024arf}, which provides a detailed treatment of subleading twist contributions in GTMDs, the GTMD correlator $W^{\nu [\Gamma]}_{[\Lambda^{N_i}\Lambda^{N_f}]}$ and the GTMD $E_{2,1}$ are functions of various kinematic variables. They can be expressed as $W^{\nu [\Gamma]}_{[\Lambda^{N_i}\Lambda^{N_f}]}(x, \xi, p_{\perp}, \Delta_{\perp}, \bfp \cdot \Dp)$ and $W^{\nu [\Gamma]}_{[\Lambda^{N_i}\Lambda^{N_f}]}(x, p_{\perp}, \Delta_{\perp}, \theta)$, respectively. 
%
%
In the above expression, $\Gamma$ represents the subleading twist Dirac matrices. In a symmetric frame, the proton's average momentum is defined as  $P = \frac{1}{2} (P^{f} + P^{i})$, while $\Delta = (P^{f} - P^{i})$ denotes the momentum transfer. 
There are a total of $32$ GTMDs for subleading twist Dirac matrices. For \(\Gamma = 1\), the GTMD correlator relates as
%
	\begin{eqnarray}
		W_{[\Lambda^{N_i}\Lambda^{N_f}]}^{[1]}
		&=& \frac{1}{2P^+} \, \bar{u}(P^{f}, \Lambda^{N_f}) \, \bigg[
		\mathbf{E_{2,1}}
		+ \frac{i\sigma^{i+} p_{\perp}^i}{P^+} \,\mathbf{E_{2,2}}
		+ \frac{i\sigma^{i+} \Delta_{\perp}^i}{P^+} \, 	\mathbf{E_{2,3}}+ \frac{i\sigma^{ij} p_{\perp}^i \Delta_{\perp}^j}{M^2} \, 	\mathbf{E_{2,4}}
		\bigg] \nonumber\\*
		&& \times~u(P^{i}, \Lambda^{N_i})
		\,. \label{par1}  
	\end{eqnarray}
In the above parameterization equation, the \textbf{bold} terms highlight the subleading twist GTMDs. All calculations adhere to the conventions outlined in Ref.~\cite{Sharma:2024arf}.
\section{Results}\label{secresults}
\noindent 
To calculate the GTMD $E_{21}^{\nu}$, we utilize the proton state representations in Eqs.~(\ref{fockSD}) and (\ref{fockVD}), along with the GTMD correlator defined in Eq.~(\ref{corr}). For specific helicity configurations, the GTMD $E_{21}^{\nu}$ can be expressed as
		\begin{eqnarray}
E_{2,1}^{\nu} &=& \frac{P^+}{2 M}\Bigg( ~W^{\nu[1]}_{[++]}+W^{\nu[1]}_{[--]}\Bigg).\ \nonumber
	\end{eqnarray}
From the above equation, the explicit expressions for \( E_{2,1}^{\nu} \) corresponding to active $u$ and $d$ quark states are given as
    \begin{eqnarray*}
	%
	%
	E_{2,1}^{u} &=&  \bigg(\frac{C_{S}^{2} N_s^2}{16 \pi^3}+\frac{C_{V}^{2}}{16 \pi^3}  \bigg(\frac{1}{3} |N_0^u|^2+\frac{2}{3}|N_1^u|^2 \bigg)\bigg)\Bigg[  m\Bigg(\frac{|\varphi_1^u|^2}{xM} +  \bigg(\bfp^2-(1-x)^2\frac{\Dp^2}{4} \bigg) \frac{|\varphi_2^u|^2}{x^3 M^3}\Bigg)\nonumber \\ && +  (1-x)\Dp^2 \frac{|\varphi_1^u||\varphi_2^u|}{2x^2 M^2}\Bigg], \label{e21u}\\
	%
	%
	E_{2,1}^{d} &=&  \frac{C_{VV}^{2}}{16 \pi^3}  \bigg(\frac{1}{3} |N_0^d|^2+\frac{2}{3}|N_1^d|^2 \bigg)\Bigg[  m\Bigg(\frac{|\varphi_1^d|^2}{xM} +  \bigg(\bfp^2-(1-x)^2\frac{\Dp^2}{4} \bigg) \frac{|\varphi_2^d|^2}{x^3 M^3}\Bigg)+  (1-x)\Dp^2 \frac{|\varphi_1^d||\varphi_2^d|}{2x^2 M^2}\Bigg]. \label{e21d}
    \end{eqnarray*}
\section{Discussion}\label{secdiscussion}
\begin{figure}
	\centering
	\begin{minipage}[c]{0.95\textwidth}
		\includegraphics[width=.45\textwidth]{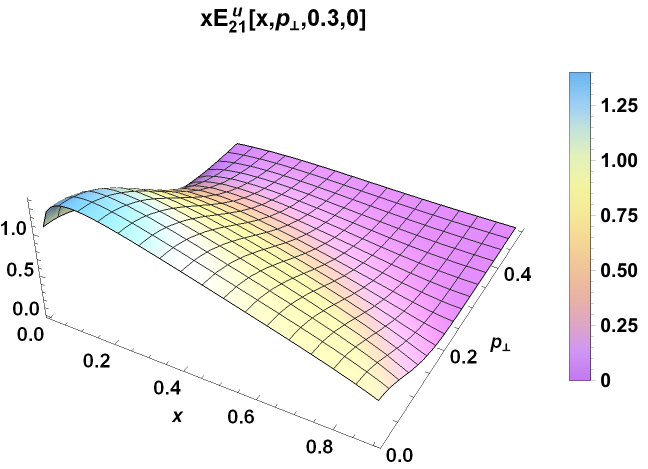}
		\includegraphics[width=.45\textwidth]{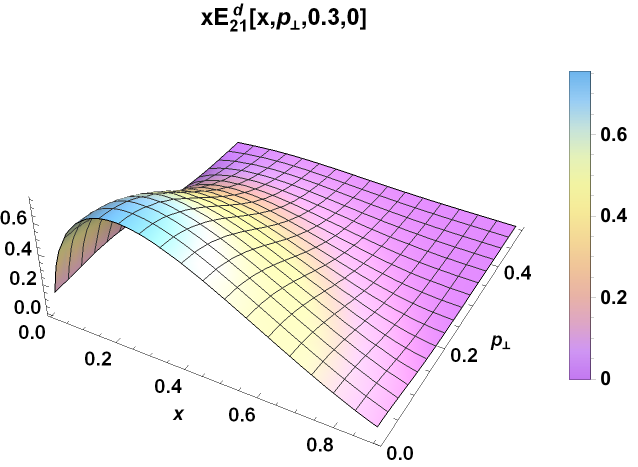}
			\caption{\label{f1} The twist-3 GTMD 		$xE_{2,1}^{\nu}$
		 plotted with respect to $x$ and ${{ p_\perp}}$ for ${ \Delta_\perp}= 0.3~\mathrm{GeV}$ for ${\bfp} \parallel {\Dp}$. The left and right plots correspond to $u$ and $d$ quarks sequentially.}
	\end{minipage}
\end{figure}

\begin{figure}
	\centering
	\begin{minipage}[c]{0.95\textwidth}
		\includegraphics[width=.45\textwidth]{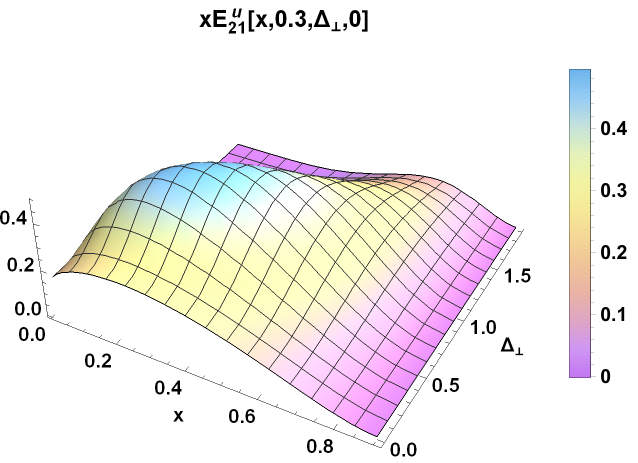}
		\includegraphics[width=.45\textwidth]{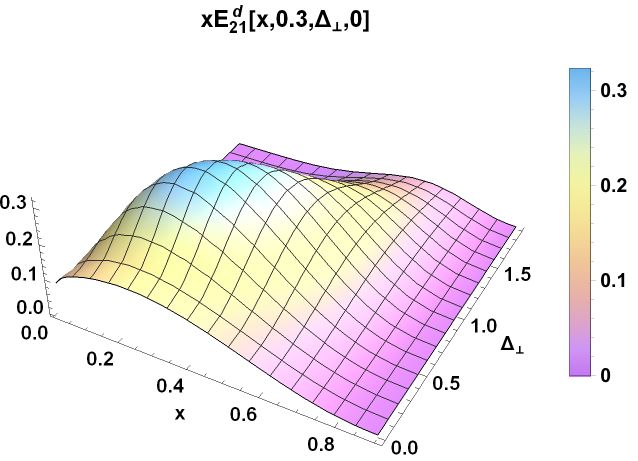}
		\caption{\label{f2} The twist-3 GTMD 		
		$xE_{2,1}^{\nu}$ plotted with respect to $x$ and ${{ \Delta_\perp}}$ for ${ p_\perp}= 0.3~\mathrm{GeV}$ for ${\bfp} \parallel {\Dp}$. The left and right plots correspond to $u$ and $d$ quarks sequentially. 
		}
	\end{minipage}
\end{figure}

\begin{figure}
	\centering
	\begin{minipage}[c]{0.95\textwidth}
		\includegraphics[width=.45\textwidth]{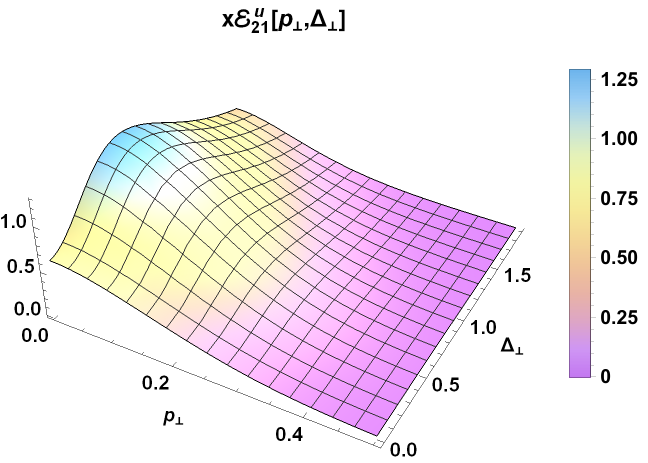}
		\includegraphics[width=.45\textwidth]{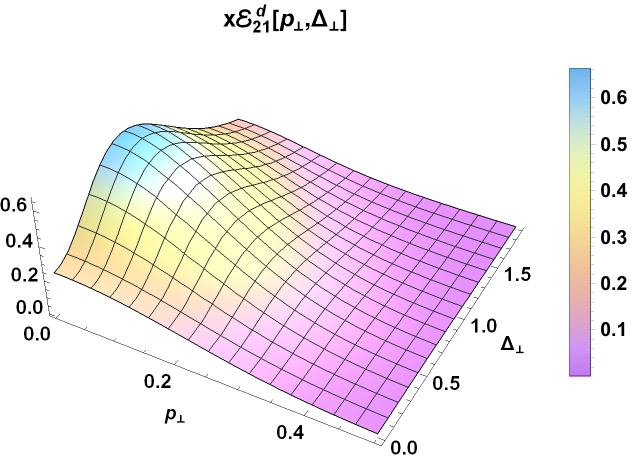}
		\caption{\label{f3} The twist-3 TMFF 		
		$x\mathcal{E}_{2,1}^{\nu}$
		plotted with respect to ${ p_\perp}$ and ${{ \Delta_\perp}}$ for ${\bfp} \parallel {\Dp}$. The left and right plots correspond to $u$ and $d$ quarks sequentially.
		}
	\end{minipage}
\end{figure}

\noindent 
In this section, we analyze the behavior of the subleading twist GTMD $xE_{2,1}$ under different kinematic configurations. Fig.~\ref{f1} illustrates $xE_{2,1}$ as a function of the longitudinal momentum fraction $x$ and the transverse momentum of the active quark $p_{\perp}$ for a fixed value of $\Delta_{\perp}$. From the plots in Fig.~\ref{f1}, it is evident that the maximum value of $xE_{2,1}$ occurs in the low-$x$ region for both $u$ and $d$ quark flavors. This indicates that, under the given kinematic configuration, quarks inside the proton are more likely to carry a low longitudinal momentum fraction.
Fig.~\ref{f2} shows the GTMD $xE_{2,1}$ as a function of $x$ and the transverse momentum transfer to the proton $\Delta_{\perp}$, while keeping $p_{\perp}$ fixed. In this case, the plots reveal that there is a non-negligible contribution from quarks with higher $\Delta_{\perp}$ values, unlike the behavior observed for $p_{\perp}$ in Fig.~\ref{f1}. 
However, the maximum contribution remains in the lower $x$ region, consistent with previous observations.
To further analyze the integrated behavior, we computed the transverse momentum form factor (TMFF) $x\mathcal{E}_{2,1}^{\nu}$ by integrating GTMD $xE_{2,1}$ over $p_{\perp}$ and $\Delta_{\perp}$
in Fig.~\ref{f3}. The plots in Fig.~\ref{f3} corroborate the findings from Fig.~\ref{f2}, highlighting contributions from configurations with higher $\Delta_{\perp}$, a feature not present for $p_{\perp}$. Importantly, the maximum contribution originates from configurations with low $p_{\perp}$ values.
%
%
%
These observations underscore the distinct kinematic dependencies of \( xE_{2,1} \) and its integrated form, shedding light on the interplay between transverse momentum and momentum transfer in the subleading twist dynamics of the proton.

\section{Conclusion}\label{seccon}
\noindent 
In this study, we investigated the subleading twist GTMD  $xE_{21}^\nu$ within the framework of the LFQDM, focusing on its behavior and physical significance. Our analysis demonstrates that this distribution exhibits quark flavor symmetry, maintaining positivity across the entire kinematic range for both $u$ and $d$ quarks. These features provide valuable constraints on the non-perturbative structure of hadrons, contributing to a deeper understanding of their underlying dynamics.
%
\par
Beyond their primary association with double Drell-Yan processes, GTMDs establish crucial connections among GPDs, TMDs, and PDFs, offering a unified framework for exploring the hadron structure. These interrelations bridge theoretical predictions with experimental observables from processes such as Deeply Virtual Compton Scattering and Semi-Inclusive Deep Inelastic Scattering, enhancing our comprehension of partonic interactions and distributions.
\par
Our findings underscore the critical role of subleading twist contributions in elucidating the intricate interplay between parton-level dynamics and the emergent properties of hadrons. By employing the LFQDM, we provide a model-based perspective that captures essential features of these distributions, laying a foundation for further theoretical investigations and experimental validations.
\par
Future research can leverage these insights to refine our understanding of hadronic tomography, particularly as next-generation experimental facilities like the EICs become operational. These efforts will be pivotal in advancing the study of hadron structure and the non-perturbative aspects of QCD.
\section{Acknowledgement}
\noindent 
H.D. would like to thank the Science and Engineering Research Board, Anusandhan-National Research Foundation, Government of India under the scheme SERB-POWER Fellowship (Ref No. SPF/2023/000116) for financial support.

\end{document}